\documentclass{IEEEtran}
\usepackage{cite}
\usepackage{bm}
\usepackage{amsmath,amssymb,amsfonts,amsthm}
\usepackage{algorithmic}
\usepackage{graphicx}
\usepackage{textcomp}
\def\BibTeX{{\rm B\kern-.05em{\sc i\kern-.025em b}\kern-.08em
    T\kern-.1667em\lower.7ex\hbox{E}\kern-.125emX}}
\begin{document}
\title{Dark Modes in Non-Markovian Linear Quantum Systems}
\author{Shikun Zhang, Daoyi Dong, \IEEEmembership{Senior Member, IEEE}, and Kun Liu, \IEEEmembership{Senior Member, IEEE}
\thanks{This work is supported by National Natural Science Foundation of China under Grant 61873034,  National Science Foundation of Beijing Municipality under Grant 4182057, the Open Subject of Beijing Intelligent Logistics System Collaborative Innovation Center under Grant BILSCIC-2019KF-13, and in part by the State Key Laboratory of Synthetical Automation for Process Industries. It is also partially supported by the Australian Research Council’s Discovery Projects funding scheme under Project DP190101566. \textit{(Corresponding author: Kun Liu.)}}
\thanks{Shikun Zhang is with School of Automation, Beijing Institute of Technology, Beijing, 100081, China (e-mail: zhang.shikun@outlook.com).}
\thanks{Daoyi Dong is with School of Engineering and Information Technology, University of New South Wales, Caberra ACT 2600, Australia (e-mail: daoyidong@gmail.com).}
\thanks{Kun Liu is with School of Automation, Beijing Institute of Technology, Beijing, 100081, China (e-mail: kunliubit@bit.edu.cn).}
}

\maketitle

\begin{abstract}
In this note, we are concerned with dark modes in a class of non-Markovian open quantum systems. Based on a microscopic model, a time-convoluted linear quantum stochastic differential equation and an output equation are derived to describe the system dynamics. The definition of dark modes is given building on the input-output structure of the system. Then, we present a necessary and sufficient condition for the existence of dark modes. Also, the problem of dark mode synthesis via Hamiltonian engineering is constructively solved and an example is presented to illustrate our results.
\end{abstract}


\section{Introduction}
An open quantum system inevitably interacts with its environment as a whole. However, certain degrees of freedom in the system may be unaffected by environmental noise, evolving in a way similar to closed quantum systems. Possible existence of decoherence-free subspaces in open systems presents an effective countermeasure against environmental effect which hinders many quantum information processing tasks.Hence, the investigation on decoherence-free subspaces has attracted wide attention in the past three decades \cite{PhysRevLett.81.2594,PhysRevA.91.042303,PhysRevA.99.062340,4639467,PhysRevA.99.042331,6777569,7870640,Dong1609,PhysRevLett.108.153603}. 

Specifically, in terms of certain continuous-variable systems, the Heisenberg evolution of some linear combinations of canonical operators, which are called ``dark modes" in literature, is immune to environmental noise. Dark modes can be conveniently characterized in Markovian linear quantum systems \cite{4625217,NURDIN20091837,7480785,LEVITT2018255,7556290,GRIVOPOULOS2018103} described by quantum stochastic differential equations (QSDEs) and have important applications in quantum technology\cite{Dong1609}. Stemming from the input-output formalism \cite{PhysRevA.31.3761}, these systems are well defined under the framework of quantum stochastic calculus \cite{1984Quantum}. In \cite{6777569}, dark modes are related to the concept of uncontrollable and unobservable linear subsystems, in accordance with the fact that such modes are decoupled from input noise and output processes. The connection built in \cite{6777569} opens the possibility of applying control theory to the analysis of dark modes in the Heisenberg picture. Moreover, a direct method for characterizing dark modes in Markovian linear quantum systems based on rank defect examination was proposed in \cite{7870640}. 

In many practical applications, the assumptions that lead to Markovian quantum dynamics do not always hold\cite{RevModPhys.89.015001}. Non-Markovian quantum systems have been widely investigated in various areas of quantum optics, quantum dot systems and quantum superconducting circuits\cite{PhysRevA.94.063848,2020Detecting,PhysRevA.101.042327,8820133} and have significant connections with unique quantum characteristics such as coherence. To the best of our knowledge, few existing works have discussed dark modes in non-Markovian open quantum systems. This note aims at presenting several results on analysis and synthesis of dark modes for a class of non-Markovian linear quantum systems. The systems can be characterized by time-convoluted linear QSDEs \cite{PhysRevA.87.032117,sync} which can be obtained by removing the assumption of uniform coupling strength with respect to different frequencies for quadratic Hamiltonians and linear coupling operators. In particular, we establish a necessary and sufficient condition for existence of dark modes and solve the synthesis problem of dark modes for this class of non-Markovian quantum systems.

The rest of this note is organized as follows. In Section II, we present the dynamics of non-Markovian linear quantum systems. Section III presents the definition of dark modes and presents a necessary and sufficient condition on existence of dark modes in the class of non-Markovian quantum systems under consideration. In Section IV, the synthesis of dark modes is investigated. An illustrative example is presented in Section V and concluding remarks are given in Section VI.

\textbf{Notation.} Roman type character i represents the imaginary unit while italic type $i$ is used for indexing. Let $A$, $B$ be arbitrary matrices and $\hat{A}$, $\hat{B}$ be Hilbert space operators. We denote by $A^T$ the transpose of $A$. $\text{Re}(A)$ and $\text{Im}(A)$ denote the real and imaginary parts of $A$, respectively. $\text{Det}(A)$ stands for the determinant of $A$. $\text{Ker}(A)$ is the kernel space of $A$, and $\text{col}(A)$ is the column space of $A$. $A \otimes B$ is the Kronecker product of $A$ and $B$. $\hat{A}^\dagger$ is the adjoint operator of $\hat{A}$. $[\hat{A},\hat{B}]=\hat{A}\hat{B}-\hat{B}\hat{A}$ denotes the commutator of $\hat{A}$ and $\hat{B}$. For positive integer $n$, $I_n$ is the $n \times n$ identity matrix. We denote the matrix $\begin{pmatrix}0&1\\-1&0\end{pmatrix}$ as $J$ and $I_n \otimes J$ as $J_n$. $\mathbb{R}^{n}$ and $\mathbb{C}^{n}$ denote real space and complex space of dimension $n$, respectively. Also, for positive integers $i$ and $j$, $\delta_{ij}$ stands for Kronecker delta function. $\delta(\cdot)$ represents Dirac delta function.

\section{System Dynamics}

Building on relevant results in \cite{PhysRevA.87.032117,sync}, we present the derivation of a non-Markovian quantum input-output system on which our study of dark modes is based directly from the microscopic model.  

Consider a localized quantum system with Hamiltonian $H_S$. The system interacts with $M$ bosonic fields modelled as an infinite number of harmonic oscillators with a continuum of frequencies. The annihilation operators of the fields are denoted as $b_1(\omega),...,b_M(\omega)$, which satisfy the following commutation relation:
\begin{equation}
[b_i(\omega),b_{j}^{\dagger}(\tilde{\omega})]=\delta_{ij}\delta(\omega-\tilde{\omega}), \quad 1\leq i,j \leq M.
\end{equation}

The bath of the system comprises of these bosonic fields, whose Hamiltonian is written as
\begin{equation}
H_B=\sum_{j=1}^{M}\int_{-\infty}^{+\infty}\omega b_{j}^{\dagger}(\omega)b_{j}(\omega) d\omega.
\end{equation}

The system-bath interaction Hamiltonian is expressed as
\begin{equation}
H_{\text{int}}=\text{i}\sum_{j=1}^{M}\int_{-\infty}^{+\infty}\Big{(}\kappa_{j}(\omega) b_{j}^{\dagger}(\omega)L_j-\kappa_{j}^{*}(\omega) b_{j}(\omega)L_j^{\dagger}\Big{)}d\omega,
\end{equation}
where $\kappa_{j}(\omega)$ ($1\leq j\leq M$) represent frequency dependent coupling strengths and are assumed to be real in this note.

In the Heisenberg picture, the evolution of a system operator $A$ obeys the following equation:
\begin{equation}
\dot{A}=-\text{i}[A, H_S+H_B+H_{\text{int}}].
\end{equation}

It can be further derived \cite{sync} that
\begin{multline}
\dot{A}=-\text{i}[A, H_S]+\bm{\tilde{b}_\text{in}}^{\dagger}[A,\bm{L}]+[\bm{L}^{\dagger},A]\bm{\tilde{b}_\text{in}}\\
+\int_0^t \Big{(}\bm{L}^{\dagger}(\tau) \Gamma^{\dagger}(t-\tau)[A,\bm{L}]+[\bm{L}^{\dagger},A]\Gamma(t-\tau)\bm{L}(\tau)\Big{)} d\tau,
\end{multline}
where $\bm{\tilde{b}_\text{in}}=(\tilde{b}_{\text{in},1},...,\tilde{b}_{\text{in},M})^T$ denotes the vector of colored noise input with 
\begin{align}
\tilde{b}_{\text{in},j}(t)=\int_{-\infty}^{+\infty}\kappa_j(\omega)e^{-\text{i}\omega t}b_j(\omega)d\omega \quad 1\leq j \leq M.
\end{align}

Also, $\Gamma(t)=\text{diag}(\gamma_1(t),...,\gamma_M(t))$ is the memory kernel matrix function where 
\begin{equation}
\gamma_j(t)=\int_{-\infty}^{+\infty}e^{-\text{i}\omega t}\kappa_j^2(\omega)d\omega,\quad 1\leq j \leq M.
\end{equation}

It can be checked that the following commutation relations hold:
\begin{equation}
[\tilde{b}_{\text{in},i}(t),\tilde{b}_{\text{in},j}^{\dagger}(\tilde{t})]=\delta_{ij}\gamma_j(t-\tilde{t}), \quad 1\leq i,j \leq M.
\end{equation}

Next, for $0 \leq t <\tilde{t}$ and $1\leq j \leq M$, we denote by $b_j(\omega, \tilde{t})$ the Heisenberg evolution of the annihilation operator at time $\tilde{t}$. For $1\leq j \leq M$, the output fields are expressed as 
\begin{equation}
\tilde{b}_{\text{out},j}(t)=\lim\limits_{\tilde{t}\to t^+}\int_{-\infty}^{+\infty}\kappa_j(\omega)e^{-\text{i}\omega (t-\tilde{t})}b_j(\omega,\tilde{t})d\omega.
\end{equation}

We present the following input-output relation:
\begin{equation}
\tilde{b}_{\text{out},j}(t)=\int_{0}^{t}\gamma_j(t-\tau)L_j(\tau)d\tau + \tilde{b}_{\text{in},j}(t).
\end{equation}
It should be noted that input-output relation (10) is different from that presented in \cite{PhysRevA.87.032117}. The output field in \cite{PhysRevA.87.032117} is still driven by white noise whereas our output field is driven by colored noise. Please see Appendix for the derivation of (10).

Then, we consider a vector of $2n$ system operators $x=(x_1,...,x_{2n})^T$ with the following canonical commutation relation (CCR):
\begin{equation}
[x,x^T]=\text{i}J_n.
\end{equation}
The system Hamiltonian and coupling operators are assumed to be quadratic and linear in $x$:
\begin{equation}
H_S=\frac{1}{2}x^T \Omega x, \quad L_j=v_j^T x,
\end{equation}
with $\Omega \in \mathbb{R}^{2n\times 2n}$, $\Omega =\Omega^T$ and $v_j \in \mathbb{C}^{2n}$ ($1\leq j \leq M$). The self-adjoint input and output vectors are introduced as
$\tilde{\omega}_{\text{in}}=(\tilde{Q}_1^{\text{in}},\tilde{P}_1^{\text{in}},...,\tilde{Q}_M^{\text{in}},\tilde{P}_M^{\text{in}})^T$ and $\tilde{\omega}_{\text{out}}=(\tilde{Q}_1^{\text{out}},\tilde{P}_1^{\text{out}},...,\tilde{Q}_M^{\text{out}},\tilde{P}_M^{\text{out}})^T$, where 
\begin{equation}
\tilde{Q}_j^{\text{in}}=\frac{\tilde{b}_{\text{in},j}+\tilde{b}_{\text{in},j}^{\dagger}}{\sqrt{2}}, \quad \tilde{P}_j^{\text{in}}=\frac{\tilde{b}_{\text{in},j}-\tilde{b}_{\text{in},j}^{\dagger}}{\sqrt{2}\text{i}},
\end{equation}
and
\begin{equation}
\tilde{Q}_j^{\text{out}}=\frac{\tilde{b}_{\text{out},j}+\tilde{b}_{\text{out},j}^{\dagger}}{\sqrt{2}}, \quad \tilde{P}_j^{\text{out}}=\frac{\tilde{b}_{\text{out},j}-\tilde{b}_{\text{out},j}^{\dagger}}{\sqrt{2}\text{i}},
\end{equation}
for $1\leq j \leq M$.

Combining (5) and (10)-(14), we obtain the following non-Markovian quantum input-output system with respect to $x$:
\begin{align}
&\dot{x}=A_H x+\int_0^t A_\Gamma (t-\tau)x(\tau)d\tau+B\tilde{\omega}_{\text{in}}\\ \nonumber
&\tilde{\omega}_{\text{out}}=\int_0^t \Gamma_o (t-\tau)Vx(\tau)d\tau+\tilde{\omega}_{\text{in}},
\end{align}
where 
\begin{align}
&A_H=J_n \Omega \\ \nonumber
&V=\sqrt{2}(\text{Re}(v_1),\text{Im}(v_1),...,\text{Re}(v_M),\text{Im}(v_M))^T \\ \nonumber
&A_\Gamma (t)=J_n V^T (\text{Im}(\Gamma(t))\otimes I_2+\text{Re}(\Gamma(t))\otimes J)V \\ \nonumber
&B=J_n V^T J_M\\ \nonumber
&\Gamma_o (t)=\text{Re}(\Gamma(t))\otimes I_2-\text{Im}(\Gamma(t))\otimes J.
\end{align}

It is clear from (16) that the non-Markovian quantum system is completely specified by $\Omega$, $V$ and $\Gamma(t)$. We make the following assumption regarding $\Gamma(t)$: 
\newcounter{assumption}
\newtheorem{assum1}[assumption]{Assumption}
\begin{assum1} 
There exists $t_1 \geq 0$ such that $\gamma_j(t_1)\neq 0$, $1\leq j \leq M$.
\end{assum1}

\section{Existence of Dark Modes}
We first give the definition of dark modes. 
\newcounter{definition}
\newtheorem{def1}[definition]{Definition}
\begin{def1} Consider $x_D=S_D x$, where $S_D \in \mathbb{R}^{2l \times 2n}$ ($l>0$) is of full row rank. $x_D$ is called a dark mode of system (15) if the following conditions are satisfied.

(C1) There exists $A_D \in \mathbb{R}^{2l \times 2l}$ such that
\begin{equation}
dx_D=A_D x_D dt,
\end{equation}
i.e., the evolution of $x_D$ induced by (15) is autonomous, time-local and decoupled from $\tilde{\omega}_{\text{in}}$.

(C2) $\tilde{\omega}_{\text{out}}$ is decoupled from $x_D$.

(C3) The following CCR is satisfied:
\begin{equation}
[x_D,x_D^T]=\emph{i}J_l. 
\end{equation}
\end{def1}

Definition 1 says that dark modes are linear combinations of specific system variable operators. On one hand, the evolution of dark modes is unperturbed by input noise. On the other hand, system output does not contain any information about dark modes. Moreover, the requirement that CCR (18) should be satisfied implies that dark modes comprise of canonical operators.

To proceed, some preliminaries on symplectic vector spaces are presented. For $\forall x,y \in \mathbb{R}^{2n}$, we define $\omega_n(x,y)\triangleq x^T J_n y$. Clearly, $\omega_n(\cdot,\cdot)$ is an anti-symmetric and non-degenerate bilinear form on $\mathbb{R}^{2n} \times \mathbb{R}^{2n}$. Therefore, $\mathbb{R}^{2n}$ is a symplectic vector space with symplectic form $\omega_n(\cdot,\cdot)$.

Let $W$ be a subspace of $\mathbb{R}^{2n}$. We denote by $W^{\tilde{\perp}}$ the symplectic orthogonal complement of $W$, i.e.,
\begin{equation*}
W^{\tilde{\perp}}=\{v\in \mathbb{R}^{2n}|\omega_n(u,v)=0, \forall u \in W\}.
\end{equation*}
If $W \cap W^{\tilde{\perp}}=\{0\}$, or equivalently if the restriction of $\omega_n(\cdot,\cdot)$ on $W$ is non-degenerate, then $W$ is called a symplectic subspace.

We now present a necessary and sufficient condition for the existence of dark modes in system (15).
\newcounter{theorem}
\newtheorem{the1}[theorem]{Theorem}
\begin{the1} 
System (15) admits dark modes if and only if there exists $W_D \subset \emph{Ker}(VJ_n)$, where $W_D$ is an $\Omega J_n$-invariant symplectic subspace of $\mathbb{R}^{2n}$ with dimension greater than 0.
\end{the1}

\begin{proof}
(Sufficiency) Suppose that $W_D$ is an $\Omega J_n$-invariant symplectic subspace of $\mathbb{R}^{2n}$ contained in $\text{Ker}(VJ_n)$ and $\text{dim}(W_D)=2l_D>0$. 

We choose a symplectic basis \cite{2019Introduction} $\{e_1,f_1,...,e_{l_D},f_{l_D}\}$ of $W_D$, i.e., for $1\leq i,j \leq l_D$
\begin{align}
\omega_n(e_i,e_i)&=0,\quad \omega_n(f_i,f_i)=0,\\ \nonumber
\omega_n(e_i,f_j)&=\delta_{ij}=-\omega_n(f_j,e_i).
\end{align}

Since $W_D$ is a symplectic subspace, $W_D^{\tilde{\perp}}$ is also a symplectic subspace. Moreover, given that $\text{dim}(W_D)+\text{dim}(W_D^{\tilde{\perp}})=2n$, we can also choose a symplectic basis $\{e_{l_D +1},f_{l_D +1},...,e_{n},f_{n}\}$ of $W_D^{\tilde{\perp}}$.

Next, the following matrices are constructed:
\begin{equation}
S_D=(e_1,f_1,...,e_{l_D},f_{l_D})^T,
\end{equation}
\begin{equation}
S_B=(e_{l_D +1},f_{l_D +1},...,e_{n},f_{n})^T.
\end{equation}
Let $x_D\triangleq S_D x$ and $x_B\triangleq S_B x$. We shall prove that $x_D$ is a dark mode.

Because the column vectors of $S_D^T$ and $S_B^T$ are symplectic bases, we have
\begin{align}
S_D J_n S_D^T&=J_{l_D}, \\
S_B J_n S_B^T&=J_{n-l_D}.
\end{align}
Eq. (22) implies that $[x_D,x_D^T]=\text{i}J_{l_D}$, thus meeting condition (C3). Also, by the definition of symplectic orthogonal complement, the following equalities hold:
\begin{align}
S_D J_n S_B^T&=0, \\
S_B J_n S_D^T&=0.
\end{align}
Let $S=(S_D^T\quad S_B^T)^T$. Eqs. (22)-(25) yield 
\begin{equation}
S J_n S^T=J_n,
\end{equation}
which means that $S$ is a symplectic matrix.

Then, we consider an equivalent system of (15) with variable $y\triangleq Sx=(x_D^T\quad x_B^T)^T$, which is written as
\begin{align}
&\dot{y}=SA_H S^{-1} y+\int_0^t SA_\Gamma (t-\tau)S^{-1}y(\tau)d\tau+SB\tilde{\omega}_{\text{in}},\\ \nonumber
&\tilde{\omega}_{\text{out}}=\int_0^t \Gamma_o (t-\tau)VS^{-1}y(\tau)d\tau+\tilde{\omega}_{\text{in}}.
\end{align}
Since $W_D \subset \text{Ker}(VJ_n)$, we have
\begin{align}
SB&=\begin{pmatrix} S_D \\ S_B \end{pmatrix} J_n V^T J_M \\ \nonumber
     &=\begin{pmatrix} 0 \\ S_B J_n V^T J_M  \end{pmatrix}.
\end{align}
Because $S$ is a symplectic matrix, we have $S^{-1}=-J_n S^T J_n$. Therefore, for $t \geq 0$
\begin{align}
\Gamma_o (t)VS^{-1}&=-\Gamma_o (t)V\begin{pmatrix}J_n S_D^T & J_n S_B^T\end{pmatrix} \begin{pmatrix}J_{l_D} & 0\\0 &J_{n-l_D} \end{pmatrix} \\ \nonumber
                                   &=-\Gamma_o (t)V\begin{pmatrix}J_n S_D^T J_{l_D}  & J_n S_B^T J_{n-l_D} \end{pmatrix}\\ \nonumber
                                   &=\begin{pmatrix}0  & -\Gamma_o (t)V J_n S_B^T J_{n-{l_D}} \end{pmatrix}.
\end{align}
Also, denoting $\Gamma_K(t)\triangleq \text{Im}(\Gamma(t))\otimes I_2+\text{Re}(\Gamma(t))\otimes J$, for $t \geq 0$, we have 
\begin{align}
S A_\Gamma (t)S^{-1}&=-\!\begin{pmatrix}S_D \\ S_B\end{pmatrix}J_n V^T \Gamma_K(t)V \begin{pmatrix}J_n S_D^T J_l  & J_n S_B^T J_{n-{l_D}} \end{pmatrix} \\\nonumber
                                    &=-\!\begin{pmatrix}0&0\\0&S_B J_n V^T \Gamma_K(t)V J_n S_B^T J_{n-{l_D}} \end{pmatrix}.
\end{align}
Since $W_D$ is $\Omega J_n$-invariant, there exists a matrix $P_1$ such that $\Omega J_n S_D^T=S_D^T P_1$. Therefore, 
\begin{equation}
S_B J_n \Omega J_n S_D^T=S_B J_n S_D^T P_1=0.
\end{equation}
Then, we proceed to show that $W_D^{\tilde{\perp}}$ is also $\Omega J_n$-invariant. For $\forall v \in W_D^{\tilde{\perp}}$ and $\forall u \in W_D$,
\begin{align}
\omega_n (u,\Omega J_n v)&=u^T J_n \Omega J_n v \\ \nonumber
                                          &=v^T J_n \Omega J_n u \\ \nonumber
                                          &=\omega_n (v,\Omega J_n u) \\ \nonumber
                                          &=0.
\end{align}
The last equality holds because $W_D$ is $\Omega J_n$-invariant. Eq. (32) shows that $\Omega J_n v \in W_D^{\tilde{\perp}}$, which indicates that $W_D^{\tilde{\perp}}$ is $\Omega J_n$-invariant. It follows that there exists a matrix $P_2$ such that $\Omega J_n S_B^T=S_B^T P_2$. We thus have
\begin{equation}
S_D J_n \Omega J_n S_B^T=S_D J_n S_B^T P_2=0.
\end{equation}
Consequently,
\begin{align}
S A_HS^{-1}&=S J_n\Omega S^{-1}\\ \nonumber
                     &=-\!\begin{pmatrix}S_D \\ S_B\end{pmatrix}J_n\Omega \begin{pmatrix}J_n S_D^T J_{l_D}  & J_n S_B^T J_{n-{l_D}} \end{pmatrix} \\ \nonumber
                     &=-\!\begin{pmatrix}S_D J_n\Omega J_n S_D^T J_{l_D} & 0\\0& S_B J_n\Omega J_n S_B^T J_{n-{l_D}}\end{pmatrix}.
\end{align}

Combining (28), (29), (30) and (34), we obtain the following equations:
\begin{align}
dx_D=-S_D J_n\Omega J_n S_D^T J_{l_D} x_D dt,
\end{align}
\begin{align}
\dot{x}_B=&-S_B J_n\Omega J_n S_B^T J_{n-{l_D}}x_B\\ \nonumber
                  &-\int_0^t S_B J_n V^T \Gamma_K(t-\tau)V J_n S_B^T J_{n-{l_D}}x_B(\tau) d\tau \\ \nonumber
                  &+S_B J_n V^T J_M \tilde{\omega}_{\text{in}},
\end{align}
\begin{align}
\tilde{\omega}_{\text{out}}=-\int_0^t \Gamma_o (t)V J_n S_B^T J_{n-{l_D}}x_B(\tau) d\tau+\tilde{\omega}_{\text{in}}.
\end{align}
Eqs. (35)-(37) indicate that conditions (C1) and (C2) are met. We have thus shown that $x_D$ is a dark mode. Moreover, it is clear that for any symplectic matrix $\tilde{S} \in \mathbb{R}^{2l_D\times 2l_D}$, $\tilde{S}x_D$ is also a dark mode.

(Necessity) Suppose that system (15) has a dark mode expressed as $\bar{x}_D=\bar{S}_D x$, where $\bar{S}_D \in \mathbb{R}^{2\bar{l}_D\times 2n}$ ($\bar{l}_D>0$). According to (C3), we have
\begin{align}
[\bar{x}_D,\bar{x}_D^T]&=\bar{S}_D[x,x^T]\bar{S}_D^T \\ \nonumber
                                        &=\text{i}\bar{S}_D J_n\bar{S}_D^T \\ \nonumber
                                        &=\text{i}J_{\bar{l}_D}.
\end{align}
We denote $\bar{S}_D^T$ as $(\bar{e}_1,\bar{f}_1,...,\bar{e}_{\bar{l}_D},\bar{f}_{\bar{l}_D})$. Eq. (38) says that, for $1\leq i,j \leq \bar{l}_D$,
\begin{align}
\omega_n(\bar{e}_i,\bar{e}_i)&=0,\quad \omega_n(\bar{f}_i,\bar{f}_i)=0,\\ \nonumber
\omega_n(\bar{e}_i,\bar{f}_j)&=\delta_{ij}=-\omega_n(\bar{f}_j,\bar{e}_i).
\end{align}
Next, for $\forall y \in \text{col}(\bar{S}_D^T)$ which is not zero, we expand $y$ as $\sum_{k=1}^{\bar{l}_D}c_k \bar{e}_k+d_k \bar{f}_k$. Then, it must be the case that there either exists $c_{k_1}\neq 0$ or $d_{k_2}\neq 0$ ($1\leq k_1,k_2 \leq \bar{l}_D$). In the former case, we have $\omega_n(y, \bar{f}_{k_1})=c_{k_1}\neq 0$, and in the latter, we have $\omega_n(y, \bar{e}_{k_2})=-d_{k_2}\neq 0$. Therefore, the restriction of symplectic form $\omega_n(\cdot, \cdot)$ on $\text{col}(\bar{S}_D^T)$ is non-degenerate, implying that $\text{col}(\bar{S}_D^T)$ is a symplectic subspace of $\mathbb{R}^{2n}$. Also, $\{\bar{e}_1,\bar{f}_1,...,\bar{e}_{\bar{l}_D},\bar{f}_{\bar{l}_D}\}$ is a symplectic basis of $\text{col}(\bar{S}_D^T)$.

We denote $\text{col}(\bar{S}_D^T)$ as $\bar{W}_D$. Then, $\bar{W}_D^{\tilde{\perp}}$ is also a symplectic subspace with symplectic basis $\{\bar{e}_{\bar{l}_D+1},\bar{f}_{\bar{l}_D+1},...,\bar{e}_{n},\bar{f}_{n}\}$. Let $\bar{S}_D^T=(\bar{e}_{\bar{l}_D+1},\bar{f}_{\bar{l}_D+1},...,\bar{e}_{n},\bar{f}_{n})$ and $\bar{S}=\begin{pmatrix} \bar{S}_D\\ \bar{S}_B \end{pmatrix}$. We now have $\bar{S}J_n\bar{S}^T=J_n$, which says that $\bar{S}$ is a symplectic matrix.

We then consider the induced dynamics of $\bar{S}x\triangleq \begin{pmatrix} \bar{x}_D\\ \bar{x}_B \end{pmatrix}$. According to (C1), $\bar{x}_D$ is decoupled from input noise $\tilde{\omega}_{\text{in}}$, which implies that $\bar{S}_D J_n V^T J_M=0$. This further indicates that $V J_n \bar{S}_D^T=0$ and that $\bar{W}_D \subset \text{Ker}(VJ_n)$. 

Next, (C1) also requires that the dynamics of $\bar{x}_D$ is decoupled from $\bar{x}_B$. Following a similar procedure to that in the proof of sufficiency, we have $\bar{S}_D J_n\Omega J_n \bar{S}_B^T J_{n-\bar{l}_D}=0$, which leads to $\bar{S}_B J_n\Omega J_n \bar{S}_D^T=0$. This says that for $\forall u \in \bar{W}_D^{\tilde{\perp}}$ and $\forall v \in \bar{W}_D$, $\omega_n(u,\Omega J_n v)=0$. As a result, $\Omega J_n v \in (\bar{W}_D^{\tilde{\perp}})^{\tilde{\perp}}=\bar{W}_D$, indicating that $\bar{W}_D$ is $\Omega J_n$-invariant.

To sum up, we have shown that there exists a symplectic, $\Omega J_n$-invariant subspace $\bar{W}_D$ contained in $\text{Ker}(VJ_n)$ with dimension $\bar{l}_D>0$. The proof is thus completed.
\end{proof}

\newcounter{remark}
\newtheorem{rem1}[remark]{Remark}
\begin{rem1} 
From Theorem 1, it is clear that $VJ_n$ cannot be of full row rank if system (15) admits dark modes. Moreover, $\text{Ker}(VJ_n)$ cannot just simply be a random nontrivial subspace of $\mathbb{R}^{2n}$. It must contain a symplectic subspace, which is in accordance with the fact that dark modes satisfy CCR. The $\Omega J_n$ invariance condition is to guarantee the autonomous evolution of dark modes.
\end{rem1}

\section{Dark Mode Synthesis via Hamiltonian Engineering}
In this section, we study how to synthesize dark modes with specified dynamics in non-Markovian quantum system (15) via Hamiltonian engineering.

There are specific quantum information processing tasks which rely on ideally closed (unitary) quantum dynamics. In the absence of environmental interactions, the unitary dynamics is pinned down by system Hamiltonian $H$. When interactions come into effect, the system dynamics is no longer unitary. However, the consequence of  environmental interactions is not always fatal. It is sometimes possible to engineer another Hamiltonian $H'$ such that a subsystem still evolves according to the unitary dynamics specified by $H$.

In terms of system (15), with $V$ and $\Gamma(t)$ fixed, we may be able to choose $\Omega$ such that the conditions in Theorem~1 is satisfied, guaranteeing the existence of dark modes. To take a step further, we are also interested in the possibility of designing specific $\Omega$ to generate dark modes with desired dynamics.

Let $E$ be an arbitrary $2n \times 2n$ matrix. We denote $V_E \triangleq VJ_n E^T$. A condition under which no dark modes can be engineered is first presented.

\newcounter{proposition}
\newtheorem{pro1}[proposition]{Proposition}
\begin{pro1} 
If $\emph{dim}(\emph{Ker}(V_S J_n)\cap\emph{Ker}(V_S))=0$ holds for all symplectic matrix $S\in \mathbb{R}^{2n\times 2n}$ , then no dark modes exist in (15).
\end{pro1}
\begin{proof}
We prove the contrapositive of Proposition 1.

Suppose that system (15) admits a dark mode with $2l_D$ operators ($l_D>0$). According to the proof of Theorem 1, there exists a symplectic matrix $S$ such that
\begin{equation}
SB=\begin{pmatrix}0\\B_2\end{pmatrix},
\end{equation}
and
\begin{equation}
\Gamma_o(t)VS^{-1}=\begin{pmatrix}0&C_2(t)\end{pmatrix},\quad \forall t \geq 0,
\end{equation}
where $B_2\in \mathbb{R}^{2(n-l_D)\times 2M}$ and $C_2(t)\in \mathbb{R}^{2M\times 2(n-l_D)}$, $t \geq 0$.

Therefore, we have
\begin{equation}
J_M V J_n S^T \begin{pmatrix}v\\0\end{pmatrix}=0,
\end{equation}
and 
\begin{equation}
-\Gamma_o(t)VJ_n S^T J_n \begin{pmatrix}v\\0\end{pmatrix}=0,\quad \forall t \geq 0,
\end{equation}
where $v\in \mathbb{R}^{2 l_D}$. Specifically, (43) holds for $t_1$ mentioned in Assumption 1. Since $\text{Det}(\Gamma_o(t_1))=\prod_{j=1}^{M}|\gamma_j(t_1)|^2 \neq 0$, $\Gamma_o(t_1)$ is invertible. As a result, for $\forall v\in \mathbb{R}^{2 l_D}$, $(v^T\quad 0)^T \in \mathbb{R}^{2n}$ is contained in $\text{Ker}(V_S J_n)\cap\text{Ker}(V_S)$. The proof is completed.
\end{proof}

Proposition 1 implies that certain coupling operators forbid the possibility of synthesizing dark modes through Hamiltonian engineering. For example, if $V$ has full row rank, it is meaningless to seek or engineer dark modes. Therefore, our discussion should be built on the premise that the condition on $V$ specified in Proposition 1 does not hold. 

In terms of a closed quantum system with canonical variable $z$ ($2k$ operators) and Hamiltonian $\frac{1}{2}z^T \hat{\Omega}z$, it can be checked that the dynamics of $z$ satisfies
\begin{equation}
dz=J_k\hat{\Omega}zdt.
\end{equation}

The following theorem states that, in terms of certain non-forbidding $V$, a wide range of dark modes can be engineered via Hamiltonian engineering whose dynamics emulates that of closed quantum systems with Hamiltonian quadratic in canonical operators.

\newtheorem{the2}[theorem]{Theorem}
\begin{the2} 
Let $V$ and $\Gamma(t)$ be fixed. Suppose that
\begin{equation*}
\emph{dim}(\emph{Ker}(V J_n)\cap\emph{Ker}(V))=2l>0. 
\end{equation*}
Then, for $\forall k, 1\leq k\leq l$ and $\forall \Omega_{\emph{Dark}}\in \mathbb{R}^{2k \times 2k}$ with $\Omega_{\emph{Dark}}=\Omega_{\emph{Dark}}^T$, there exists a symmetric matrix $\Omega \in \mathbb{R}^{2n \times 2n}$ such that the non-Markovian quantum system specified by $\Omega$, $V$ and $\Gamma(t)$ contains a dark mode $x_D$ with $2k$ operators. The dynamics of $x_D$ satisfies
\begin{equation}
dx_D=J_k \Omega_{\emph{Dark}} x_Ddt.
\end{equation}
\end{the2}
\begin{proof}
To begin with, $\text{Ker}(V J_n)\cap\text{Ker}(V)\triangleq \mathcal{H}_D$ must be of even dimension. Following similar procedures in \cite{6777569} and \cite{7870640}, an orthonormal basis of $\mathcal{H}_D$ can be constructed. We first choose a unit length vector $v_1 \in\mathcal{H}_D$. Hence, unit length vector $J_n v_1$ is orthogonal to $v_1$ and is also contained in $\mathcal{H}_D$. Next, we choose a unit length vector $v_2\in \mathcal{H}_D$ which is orthogonal to $v_1$ and $J_n v_1$. It can be verified that $v_1$, $J_n v_1$, $v_2$ and $J_n v_2$ are mutually orthogonal. This process can be repeated until we have obtained $v_l, J_n v_l$ and no more vectors orthogonal to the previously chosen ones can be obtained. Therefore, $\{v_1, J_n v_1,...,v_l, J_n v_l \}$ is an orthonormal basis of $\mathcal{H}_D$.

Suppose that $\Omega_{\text{Dark}}\in \mathbb{R}^{2k_D \times 2k_D}$, $1\leq k_D\leq l$. Since $\Omega_{\text{Dark}}$ is symmetric, we can perform the following spectral decomposition:
\begin{equation}
\Omega_{\text{Dark}}=\sum_{j=1}^{2k_D}\lambda_j \beta_j \beta_j^T,
\end{equation}
where $\{\lambda_j\}_{j=1}^{2k_D}$ are eigenvalues of $\Omega_{\text{Dark}}$ and $\{\beta_j\}_{j=1}^{2k_D}\subset \mathbb{R}^{2k_D}$ are orthonormal eigenvectors of $\Omega_{\text{Dark}}$.

Let $S_D=(J_n v_1, v_1,..., J_n v_{k_D}, v_{k_D})^T$. We make the following construction:
\begin{equation}
\Omega=\sum_{j=1}^{2k_D}\lambda_j S_D^T\beta_j \beta_j^T S_D+\sum_{j=2k_D+1}^{2n}\mu_j \alpha_j \alpha_j^T,
\end{equation}
where $\{\mu_j\}_{j=2k_D+1}^{2n} \subset  \mathbb{R}$ can be chosen arbitrarily and $\{\alpha_j\}_{j=2k_D+1}^{2n}\subset \mathbb{R}^{2n}$ are mutually orthogonal vectors in the orthogonal complement of $\text{col}(S_D^T)$.

It can be verified that $S_D S_D^T=I_{2k_D}$. For $1\leq j \leq 2k_D$, we denote $S_D^T \beta_j\triangleq \hat{\beta}_j$. Then, for $1\leq j,k \leq 2k_D$, we have 
\begin{equation}
\hat{\beta}_j^T \hat{\beta}_k=\beta_j^T S_D S_D^T \beta_k=\beta_j^T\beta_k=\delta_{jk}.
\end{equation}
Therefore, $\{\hat{\beta}_1,..,\hat{\beta}_{2k_D}\}$ is an orthonormal basis of $\text{col}(S_D^T)$.

It can also be checked that $S_D J_n S_D^T=J_{k_D}$. According to the proof of necessity for Theorem 1, $W_D\triangleq \text{col}(S_D^T)$ is a symplectic subspace of $\mathbb{R}^{2n}$. On one hand, $W_D$ is clearly $J_n$-invariant by construction. On the other hand, for $\forall y \in W_D$, we have $y=\sum_{j=1}^{2k_D}c_j \hat{\beta}_j$ and $\Omega y=\sum_{j=1}^{2k_D}\lambda_j c_j \hat{\beta}_j \in W_D$. $W_D$ is thus also $\Omega$-invariant. As a result, $W_D$ is $\Omega J_n$-invariant. Also, we have
\begin{equation*}
W_D \subset \mathcal{H}_D \subset \text{Ker}(VJ_n).
\end{equation*}
It has been shown that $W_D$ is an $\Omega J_n$-invariant symplectic subspace with dimension $2k_D$ contained in $\text{Ker}(VJ_n)$. According to Theorem 1, the system admits a dark mode, which is $x_D \triangleq S_D x$.

Following the same procedure mentioned at the beginning of this proof, we obtain an orthomormal basis of $W_D^\perp$, which is $\{v_{k_D+1}, J_n v_{k_D+1},...,v_n, J_n v_n \}$. Letting $S_B=(J_n v_{k_D+1},v_{k_D+1},...,J_n v_n,v_n)^T$, and $S=\begin{pmatrix}S_D\\S_B\end{pmatrix}$, we have $S^{-1}=S^T$ by construction. It can also be verified that
\begin{equation}
S_D J_n=J_{k_D}S_D.
\end{equation}
Therefore, we obtain the dynamics of $x_D$:
\begin{align}
dx_D&=S_D J_n \Omega S_D^T x_D dt \\\nonumber
         &=J_{k_D}S_D\Omega S_D^T x_D dt \\\nonumber
         &=J_{k_D}\sum_{j=1}^{2k_D}\lambda_j S_D S_D^T \beta_j \beta_j^T S_D S_D^T x_D dt\\\nonumber
         &=J_{k_D}\sum_{j=1}^{2k_D}\lambda_j \beta_j \beta_j^T x_D dt\\\nonumber
         &=J_{k_D}\Omega_{\text{Dark}} x_Ddt.
\end{align}
\end{proof}

\newtheorem{rem2}[remark]{Remark}
\begin{rem2} 
Theorem 2 says that, if $\emph{dim}(\mathcal{H}_D)=2l>0$, the dynamics of all closed linear quantum systems consisting of $2l$ (or less) canonical operators can be emulated by a dark mode synthesized via Hamiltonian engineering. The Hamiltonian that leads to desired dynamics can be engineered according to (46) and (47).
\end{rem2}

\section{Illustrative Example}
In this section, we present an example of dark mode synthesis in a non-Markovian open quantum system. Consider a three-mode non-Markovian quantum system with canonical variable $x=(x_1,p_1,x_2,p_2,x_3,p_3)^T$, where
\begin{align}
&[x_j,x_k]=[p_j,p_k]=0, \\ \nonumber
&[x_j,p_k]=\text{i}\delta_{jk}, \quad 1\leq j,k \leq 3.
\end{align}

The system interacts with the environment via two coupling operators: $L_1=a_1+a_2$ and $L_2=a_2+a_3$, where $a_j=\frac{1}{\sqrt{2}}(x_j+\text{i}p_j)$ are annihilation operators ($1\leq j \leq 3$). The memory kernel matrix is expressed as
\begin{equation*}
\Gamma(t)=\begin{pmatrix}\gamma_1(t)&0\\0&\gamma_2(t)\end{pmatrix},
\end{equation*}
where $\gamma_1(t)$ and $\gamma_2(t)$ are assumed to be real functions on $[0,+\infty)$.

Suppose that the system Hamiltonian can be engineered. We expect to choose a proper $\Omega$ so as to synthesize a dark mode $x_D$ with dynamics:
\begin{equation*}
dx_D=J\Omega_{\text{Dark}} x_Ddt,
\end{equation*}
where 
\begin{equation*}
\Omega_{\text{Dark}}=\begin{pmatrix}m\omega^2&0\\0&\frac{1}{m}\end{pmatrix}, m>0,\omega >0.
\end{equation*}
This corresponds to a closed quantum system with Hamiltonian 
\begin{equation*}
H_{\text{Dark}}=\frac{1}{2}x_D^T\Omega_{\text{Dark}}x_D=\frac{1}{2}m\omega^2 q_D^2+\frac{1}{2m}p_D^2,
\end{equation*}
where $[q_D,p_D]=\text{i}$.

Next, direct calculation yields
\begin{equation*}
V=\begin{pmatrix}1&0&1&0&0&0\\
                              0&1&0&1&0&0\\
                              0&0&1&0&1&0\\
                              0&0&0&1&0&1
\end{pmatrix}.
\end{equation*}
It can be further checked that 
\begin{align*}
\mathcal{H}_D&\triangleq \text{Ker}(V J_n)\cap\text{Ker}(V) \\ \nonumber
                         &= \text{span}\bigg{\{}\frac{1}{\sqrt{3}}u_1,\frac{1}{\sqrt{3}}u_2\bigg{\}},
\end{align*}
where $u_1=(1,0,-1,0,1,0)^T$ and $u_2=(0,1,0,-1,0,1)^T$.
According to Theorem 2 and Theorem 1, the desired dark mode can be synthesized via Hamiltonian engineering.

Following the procedure outlined in the proof of Theorem~2, we construct a symplectic and orthogonal matrix $S$. We have $S^T=(S_D^T \quad S_B^T)$, where
\begin{equation*}
S_D^T=\begin{pmatrix}
                              0&\frac{1}{\sqrt{3}}\\
                              -\frac{1}{\sqrt{3}}&0\\
                              0&-\frac{1}{\sqrt{3}}\\
                              \frac{1}{\sqrt{3}}&0 \\
                              0&\frac{1}{\sqrt{3}}\\
                              -\frac{1}{\sqrt{3}}&0
             \end{pmatrix},
\end{equation*}
and
\begin{equation*}
S_B^T=\begin{pmatrix}
                              \frac{1}{\sqrt{6}}&0&-\frac{1}{\sqrt{2}}&0\\
                              0&\frac{1}{\sqrt{6}}&0&-\frac{1}{\sqrt{2}}\\
                               \frac{2}{\sqrt{6}}&0&0&0\\
                              0& \frac{2}{\sqrt{6}}&0&0 \\
                              \frac{1}{\sqrt{6}}&0&\frac{1}{\sqrt{2}}&0\\
                              0& \frac{1}{\sqrt{6}}&0& \frac{1}{\sqrt{2}}
             \end{pmatrix}.
\end{equation*}
The following spectral decomposition can be obtained:
\begin{equation*}
\Omega_{\text{Dark}}=m\omega^2\begin{pmatrix}1\\0\end{pmatrix}\begin{pmatrix}1&0\end{pmatrix}+ \frac{1}{m}\begin{pmatrix}0\\1\end{pmatrix}\begin{pmatrix}0&1\end{pmatrix}.
\end{equation*}
We then follow (47) and obtain
\begin{align*}
\Omega&=m\omega^2S_D^T\begin{pmatrix}1\\0\end{pmatrix}\begin{pmatrix}1&0\end{pmatrix}S_D+ \frac{1}{m}S_D^T\begin{pmatrix}0\\1\end{pmatrix}\begin{pmatrix}0&1\end{pmatrix}S_D \\ \nonumber
            &=\begin{pmatrix}\frac{1}{3m}&0&-\frac{1}{3m}&0&\frac{1}{3m}&0\\
                                          0&\frac{m\omega^2}{3}&0&-\frac{m\omega^2}{3}&0&\frac{m\omega^2}{3}\\
                                          -\frac{1}{3m}&0&\frac{1}{3m}&0&-\frac{1}{3m}&0\\
                                          0&-\frac{m\omega^2}{3}&0&\frac{m\omega^2}{3}&0&-\frac{m\omega^2}{3}\\
                                          \frac{1}{3m}&0&-\frac{1}{3m}&0&\frac{1}{3m}&0\\
                                          0&\frac{m\omega^2}{3}&0&-\frac{m\omega^2}{3}&0&\frac{m\omega^2}{3}\\
\end{pmatrix}.
\end{align*}
The dynamics of $Sx=\begin{pmatrix}S_Dx\\S_Bx\end{pmatrix}\triangleq\begin{pmatrix}x_D\\x_B\end{pmatrix}$ is shown by direct calculation:
\begin{align}
SA_HS^{-1}&=\begin{pmatrix}A_H^{\text{Dark}}&0\\0&0\end{pmatrix}, \\ \nonumber
                SB&=\begin{pmatrix}0\\B_2\end{pmatrix}, \\ \nonumber
\Gamma_o (t)VS^{-1}&=\begin{pmatrix}0&\Gamma_o (t)V_2\end{pmatrix}, t\geq 0,\\ \nonumber
SA_{\Gamma}(t)S^{-1}&=\begin{pmatrix}0&0\\0&A_{\Gamma}^B(t)\end{pmatrix}, t\geq 0, \\ \nonumber
\end{align}
where 
\begin{align}
A_H^{\text{Dark}}&=\begin{pmatrix}0&\frac{1}{m}\\-m\omega^2&0\end{pmatrix},\\ \nonumber
B_2&=\begin{pmatrix}-\frac{3}{\sqrt{6}}&0&-\frac{3}{\sqrt{6}}&0\\
                                    0&-\frac{3}{\sqrt{6}}&0&-\frac{3}{\sqrt{6}}\\
                                    \frac{1}{\sqrt{2}}&0&-\frac{1}{\sqrt{2}}&0\\
                                    0&\frac{1}{\sqrt{2}}&0&-\frac{1}{\sqrt{2}}\end{pmatrix},\\ \nonumber
V_2&=\begin{pmatrix}\frac{3}{\sqrt{6}}&0&-\frac{1}{\sqrt{2}}&0\\
                                    0&\frac{3}{\sqrt{6}}&0&-\frac{1}{\sqrt{2}}\\
                                    \frac{3}{\sqrt{6}}&0&\frac{1}{\sqrt{2}}&0\\
                                    0&\frac{3}{\sqrt{6}}&0&\frac{1}{\sqrt{2}}\end{pmatrix},\\ \nonumber
A_{\Gamma}^B(t)&=\begin{pmatrix}f_1(t)&0&f_2(t)&0\\
                                    0&f_1(t)&0&f_2(t)\\
                                    f_2(t)&0&f_3(t)&0\\
                                    0&f_2(t)&0&f_3(t)\end{pmatrix}, t\geq 0,\\ \nonumber
f_1(t)&=-\frac{3}{2}(\gamma_1(t)+\gamma_2(t)), t\geq 0, \\ \nonumber
f_2(t)&=\frac{\sqrt{3}}{2}(\gamma_1(t)-\gamma_2(t)), t\geq 0, \\ \nonumber
f_3(t)&=-\frac{1}{2}(\gamma_1(t)+\gamma_2(t)), t\geq 0. 
\end{align}
Therefore, it is now clear that $x_D$ is a dark mode with dynamics:
\begin{equation*}
dx_D=A_H^{\text{Dark}}x_Ddt=J\Omega_{\text{Dark}}x_Ddt.
\end{equation*}

\section{Conclusion}
We have established a necessary and sufficient condition under which dark modes exist in a class of non-Markovian open quantum systems. Moreover, we have proposed a Hamiltonian engineering method for synthesizing dark modes with desired dynamics. Our results may be useful in quantum information processing tasks in realistic scenarios involving non-Markovian effects.

\section*{Appendix}
The derivation of (10) is shown here. For $t\geq 0$ and $1\leq j \leq M$, we denote by $b_j(\omega,t)$ the Heisenberg evolved annihilation operator at time $t$ with initial condition $b_j(\omega,0)=b_j(\omega)$. For $1\leq j \leq M$, the evolution of $b_j(\omega,t)$ satisfies
\begin{align}
\dot{b_j}(\omega,t)&=-\text{i}[b_j(\omega,t),H_S+H_B+H_\text{int}] \\ \nonumber
                              &=-\text{i}\omega b_j(\omega,t)+\kappa_j(\omega)L_j.
\end{align}
Therefore, on one hand, we have
\begin{equation}
b_j(\omega,t)=e^{-\text{i}\omega t}b_j(\omega,0)+\kappa_j(\omega)\int_0^t e^{-\text{i}\omega (t-\tau)}L_j(\tau)d\tau.
\end{equation}

On the other hand, the solution of $b_j(\omega,t)$ can be obtained by back propagation from $b_j(\omega,\tilde{t})$, $\tilde{t}>t$:
\begin{equation}
b_j(\omega,t)=e^{-\text{i}\omega (t-\tilde{t})}b_j(\omega,\tilde{t})+\kappa_j(\omega)\int_{\tilde{t}}^t e^{-\text{i}\omega (t-\tau)}L_j(\tau)d\tau.
\end{equation}

Eqs. (55) and (56) yield
\begin{equation}
e^{-\text{i}\omega (t-\tilde{t})}b_j(\omega,\tilde{t})\!=\!e^{-\text{i}\omega t}b_j(\omega,0)+\kappa_j(\omega)\!\int_{0}^{\tilde{t}} e^{-\text{i}\omega (t-\tau)}L_j(\tau)d\tau.
\end{equation}

We define:
\begin{equation*}
 \tilde{b}_{\text{out},j}(t,\tilde{t})\triangleq \int_{-\infty}^{+\infty}\kappa_j(\omega)e^{-\text{i}\omega (t-\tilde{t})}b_j(\omega,\tilde{t})d\omega. 
\end{equation*}
Then, we have 

\begin{equation}
\begin{aligned}
\tilde{b}_{\text{out},j}(t,\tilde{t})&=\tilde{b}_{\text{in},j}(t)+\int_{-\infty}^{+\infty}\kappa_j^2(\omega)\int_{0}^{\tilde{t}}e^{-\text{i}\omega (t-\tau)}L_j(\tau)d\tau d\omega \\ 
                                                      &=\tilde{b}_{\text{in},j}(t)+\int_{0}^{\tilde{t}}\int_{-\infty}^{+\infty}\kappa_j^2(\omega)e^{-\text{i}\omega (t-\tau)}d\omega L_j(\tau)d\tau \\ 
                                                      &=\tilde{b}_{\text{in},j}(t)+\int_{0}^{\tilde{t}}\gamma_j(t-\tau)L_j(\tau)d\tau.
\end{aligned}
\end{equation}

Therefore,
\begin{align}
\tilde{b}_{\text{out},j}(t)&=\lim\limits_{\tilde{t}\to t^+}\tilde{b}_{\text{out},j}(t,\tilde{t}) \\ \nonumber
                                         &=\tilde{b}_{\text{in},j}(t)+\int_{0}^{t}\gamma_j(t-\tau)L_j(\tau)d\tau,
\end{align}
which yields input-output relation (10).



\bibliographystyle{unsrt}        
\bibliography{reference} 

\begin{thebibliography}{10}

\bibitem{PhysRevLett.81.2594}
D.~A. {Lidar}, I.~L. {Chuang}, and K.~B. {Whaley}.
\newblock Decoherence-free subspaces for quantum computation.
\newblock {\em Phys. Rev. Lett.}, 81:2594--2597, Sep 1998.

\bibitem{PhysRevA.91.042303}
W.~{Qin}, C.~{Wang}, and X.~{Zhang}.
\newblock Protected quantum-state transfer in decoherence-free subspaces.
\newblock {\em Phys. Rev. A}, 91:042303, Apr 2015.

\bibitem{PhysRevA.99.062340}
J.~{Kattem\"olle} and J.~van {Wezel}.
\newblock Dynamical fidelity susceptibility of decoherence-free subspaces.
\newblock {\em Phys. Rev. A}, 99:062340, Jun 2019.

\bibitem{4639467}
F.~{Ticozzi} and L.~{Viola}.
\newblock Quantum {M}arkovian subsystems: Invariance, attractivity, and
  control.
\newblock {\em IEEE Transactions on Automatic Control}, 53(9):2048--2063, 2008.

\bibitem{PhysRevA.99.042331}
J.~van {Meter} and E.~{Knill}.
\newblock Approximate exchange-only entangling gates for the three-spin-$1/2$
  decoherence-free subsystem.
\newblock {\em Phys. Rev. A}, 99:042331, Apr 2019.

\bibitem{6777569}
N.~{Yamamoto}.
\newblock Decoherence-free linear quantum subsystems.
\newblock {\em IEEE Transactions on Automatic Control}, 59(7):1845--1857, 2014.

\bibitem{7870640}
Y.~{Pan}, D.~{Dong}, and I.~R. {Petersen}.
\newblock Dark modes of quantum linear systems.
\newblock {\em IEEE Transactions on Automatic Control}, 62(8):4180--4186, 2017.

\bibitem{Dong1609}
C.~{Dong}, V.~{Fiore}, M.~C. {Kuzyk}, and H.~{Wang}.
\newblock Optomechanical dark mode.
\newblock {\em Science}, 338(6114):1609--1613, 2012.

\bibitem{PhysRevLett.108.153603}
Y.~{Wang} and A.~A. {Clerk}.
\newblock Using interference for high fidelity quantum state transfer in
  optomechanics.
\newblock {\em Phys. Rev. Lett.}, 108:153603, Apr 2012.

\bibitem{4625217}
M.~R. {James}, H.~I. {Nurdin}, and I.~R. {Petersen}.
\newblock ${H}^{\infty}$ control of linear quantum stochastic systems.
\newblock {\em IEEE Transactions on Automatic Control}, 53(8):1787--1803, 2008.

\bibitem{NURDIN20091837}
H.~I. {Nurdin}, M.~R. {James}, and I.~R. {Petersen}.
\newblock Coherent quantum {LQG} control.
\newblock {\em Automatica}, 45(8):1837--1846, 2009.

\bibitem{7480785}
S.~L. {Vuglar} and I.~R. {Petersen}.
\newblock Quantum noises, physical realizability and coherent quantum feedback
  control.
\newblock {\em IEEE Transactions on Automatic Control}, 62(2):998--1003, 2017.

\bibitem{LEVITT2018255}
M.~{Levitt}, M.~{Guţă}, and H.~I. {Nurdin}.
\newblock Power spectrum identification for quantum linear systems.
\newblock {\em Automatica}, 90:255--262, 2018.

\bibitem{7556290}
S.~{Wang} and D.~{Dong}.
\newblock Fault-tolerant control of linear quantum stochastic systems.
\newblock {\em IEEE Transactions on Automatic Control}, 62(6):2929--2935, 2017.

\bibitem{GRIVOPOULOS2018103}
S.~{Grivopoulos} and I.~R. {Petersen}.
\newblock Bilinear hamiltonian interactions between linear quantum systems via
  feedback.
\newblock {\em Automatica}, 89:103--110, 2018.

\bibitem{PhysRevA.31.3761}
C.~W. {Gardiner} and M.~J. {Collett}.
\newblock Input and output in damped quantum systems: Quantum stochastic
  differential equations and the master equation.
\newblock {\em Phys. Rev. A}, 31:3761--3774, Jun 1985.

\bibitem{1984Quantum}
R.~L. {Hudson} and K.~R. {Parthasarathy}.
\newblock Quantum {Ito's} formula and stochastic evolutions.
\newblock {\em Communications in Mathematical Physics}, 93(3):301--323, 1984.

\bibitem{RevModPhys.89.015001}
I.~{de Vega} and D.~{Alonso}.
\newblock Dynamics of {non-Markovian} open quantum systems.
\newblock {\em Rev. Mod. Phys.}, 89:015001, Jan 2017.

\bibitem{PhysRevA.94.063848}
M.~{Malekakhlagh}, A.~{Petrescu}, and H.~E. {T\"ureci}.
\newblock Non-markovian dynamics of a superconducting qubit in an open
  multimode resonator.
\newblock {\em Phys. Rev. A}, 94:063848, Dec 2016.

\bibitem{2020Detecting}
K.~{Wu}, Z.~{Hou}, G.~{Xiang}, C.~{Li}, and F.~{Nori}.
\newblock Detecting non-{Markovianity} via quantified coherence: theory and
  experiments.
\newblock {\em npj Quantum Information}, 6(1):55, 2020.

\bibitem{PhysRevA.101.042327}
S.~{Zhang}, K.~{Liu}, D.~{Dong}, X.~{Feng}, and F.~{Pan}.
\newblock Subspace stabilization analysis for a class of non-{Markovian} open
  quantum systems.
\newblock {\em Phys. Rev. A}, 101:042327, Apr 2020.

\bibitem{8820133}
S.~{Xue}, T.~{Nguyen}, M.~R. {James}, A.~{Shabani}, V.~{Ugrinovskii}, and I.~R.
  {Petersen}.
\newblock Modeling for non-{Markovian} quantum systems.
\newblock {\em IEEE Transactions on Control Systems Technology},
  28(6):2564--2571, 2020.

\bibitem{PhysRevA.87.032117}
J.~{Zhang}, Y.~{Liu}, R.~{Wu}, K.~{Jacobs}, and F.~{Nori}.
\newblock Non-markovian quantum input-output networks.
\newblock {\em Phys. Rev. A}, 87:032117, Mar 2013.

\bibitem{sync}
S.~{Zhang}, K.~{Liu}, D.~{Dong}, X.~{Feng}, and F.~{Pan}.
\newblock Expectation synchronization synthesis in non-{M}arkovian open quantum
  systems.
\newblock {\em arXiv:2101.00833v2}, Feb 2021.

\bibitem{2019Introduction}
J.~L. {Koszul} and Y.~{Zou}.
\newblock {\em Introduction to Symplectic Geometry}.
\newblock Springer, Singapore, 2019.

\end{thebibliography}

\end{document}